\documentclass[openacc]{rstransa}

\usepackage[square,numbers]{natbib}

\newcommand{\cref}[1]{#1} 

\titlehead{Research}

\begin{document}

\title{Probing \cref{the properties of active regions in the solar interface region using} full-disk spectroheliograms}

\author{
\'{E}.~Power$^{1}$, D.~M.~Long$^{1,2}$, T.~Mihailescu$^{3,4}$, and L.~A.~Hayes$^{2}$}

\address{$^{1}$CfAR, School of Physical Sciences, DCU, Ireland\\
$^{2}$Astronomy \& Astrophysics Section, DIAS, Ireland\\
$^{3}$NASA GSFC, Greenbelt, MD 20771, USA\\
$^{4}$USRA, Washington DC, 20024, USA}

\subject{astrophysics, solar physics}

\keywords{Elemental Composition, Solar Chromosphere}

\corres{David Long\\
\email{david.long@dcu.ie}}

\begin{abstract}
The composition of plasma in the solar corona is characterised by the First Ionisation Potential (FIP) bias, and is thought to be the result of a ponderomotive force acting in the chromosphere to separate ionised from neutral plasma. \cref{I}dentifying potential signatures of this process \cref{in} the solar chromosphere \cref{is the subject of active research. Full disk spectroheliograms of the chromosphere and transition region from the Interface Region Imaging Spectrometer (IRIS) spacecraft provide an opportunity to compare plasma signatures between active regions at different evolutionary stages and assess their relationship with the fractionation processes.} \cref{Here} we compare the C~II, Si~IV, and Mg~II lines observed by IRIS, finding no clear variability between \cref{active regions at different evolutionary stages} in the C~II and Si~IV lines. However, \cref{distinct differences can be identified between the active regions using the Mg~II~k/h ratio (which provides a proxy for plasma opacity). In particular, the regions with} the highest median FIP bias \cref{exhibit} double peaked distributions of plasma opacity\cref{, suggesting variable plasma density which could affect wave propagation in these locations}. These results \cref{indicate} that the relationship between \cref{the plasma properties} and \cref{how the plasma is fractionated should} be investigated in more detail \cref{by combining observations and modelling to better understand how it changes on both temporal and spatial scales}.
\end{abstract}
\maketitle


\section{Introduction}

Spectroscopic observations of the solar atmosphere have revealed distinct differences between the elemental abundance of the solar photosphere and the solar corona. In particular, elements with a low ($<$10~eV) first ionisation potential (FIP) tend to be overabundant in the corona relative to the photosphere, while elements with a high ($\geq$10~eV) FIP maintain their abundance in the corona relative to the photosphere. This measured abundance ratio is defined as the FIP bias \cite{Asplund:2009,Meyer:1985}, and is an important diagnostic which has implications for physical processes such as the rate of heating and cooling of the plasma in the solar atmosphere in general as well as during distinct events such as flares \cite[cf.][]{Cook:1989,Reep:2025}. As a result, understanding the fractionation process and identifying where in the atmosphere it occurs has significant implications for our comprehension of the Sun and its effect on the heliosphere.

Current theories to explain solar FIP bias rely on the fractionation of ionised plasma by a ponderomotive force induced by resonant and non-resonant waves trapped within coronal loops and reflecting/refracting as they approach the steep density gradient of the chromosphere \cite[cf.][]{Laming:2004,Laming:2015,Murabito:2021,Murabito:2024}. In this interpretation, neutral particles with a low FIP are preferentially ionised, with the ponderomotive force then acting on the ions and transporting them into the corona. In contrast, high FIP elements remain largely neutral and hence unfractionated, producing the observed discrepancy. 

Recent studies have made significant progress in identifying spatial and temporal trends in FIP bias across different regions in the solar atmosphere. This step change has been driven primarily by the spatially resolved spectroscopic observations provided by the Extreme ultraviolet Imaging Spectrometer \cite[EIS;][]{Culhane:2007} onboard the \emph{Hinode} spacecraft \cite{Kosugi:2007} since its launch in 2006. Building on initial work by \cite{Feldman:2009}, who identified spectral line pairs observed by EIS which could be used to estimate FIP bias, \cite{Brooks:2011} developed a Markov-Chain Monte Carlo technique which they used to produce spatially resolved FIP bias maps using the Si~X~258.37~\AA/S~X~264.22~\AA\ line ratio. This technique was subsequently used by \cite{Brooks:2015} to produce a full-disk FIP bias map, thus enabling a direct comparison between active regions of different evolutionary stages.

The approach pioneered by \cite{Brooks:2011,Brooks:2015} has led to a revolution in our understanding of the evolution of FIP bias in different regions of the solar atmosphere, being used to probe e.g., eruptive flux ropes \cite{Baker:2021}, solar flares \cite{To:2021,To:2024}, and emerging and evolving active regions \cite{Baker:2013,Baker:2018,Mihailescu:2022} among other phenomena. While most of this work has been focussed on individual flares and active regions, \cite{Mihailescu:2022} used a series of full-disk mosaics produced by \emph{Hinode}/EIS to examine how FIP bias changes with active region evolution. This large-scale statistical approach found that FIP bias generally decreases as active regions decay, in contrast to the original suggestion of \cite{Widing:2001}, but consistent with the more recent work of \cite{Baker:2015}. In particular, \cite{Mihailescu:2022} found no strong correlation with magnetic flux or age, but noted weak trends related to flux density and evolutionary phase, suggesting that the mechanisms behind fractionation could vary from region to region and evolve over time.

\begin{figure}
    \centering
    \includegraphics[width=1\linewidth]{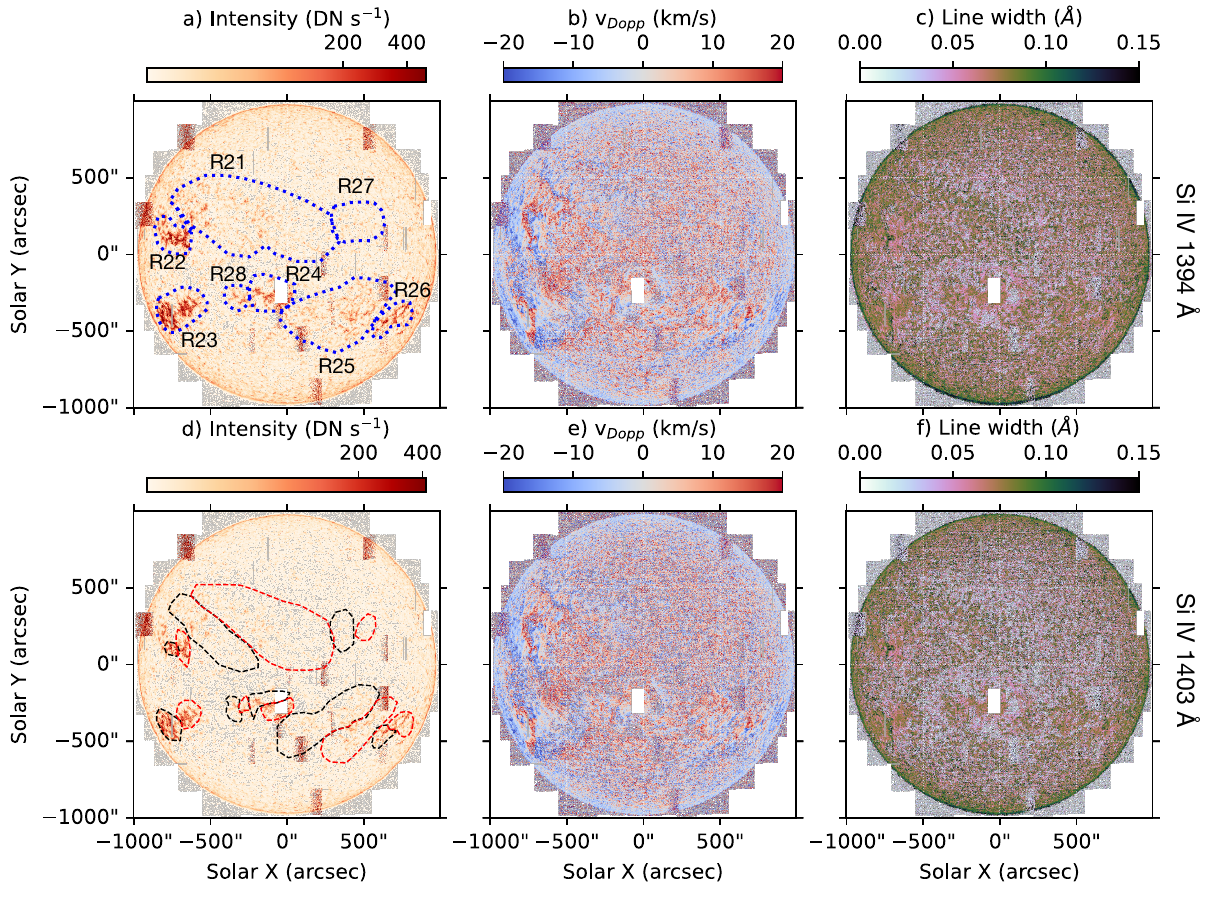}
    \caption{Full disk mosaics taken by the IRIS spacecraft in the \cref{Si~IV~1394~\AA\ (top row) and Si~IV~1403~\AA\ (bottom row)} spectral lines. Each spectral line was fitted using a single Gaussian, with the left column showing the fitted peak intensity, the centre column the Doppler velocity, and the right column the line width. Panel~a shows the active regions defined by \cite{Mihailescu:2022} and analysed here, with \cref{panel~d} showing the leading (red) and following (black) polarities of the corresponding active regions.}
    \label{fig:siiv_context}
\end{figure}

Although these results provide key insights into the measureable and observable coronal manifestation of the FIP effect, the details of the process driving this fractionation remain elusive. The ponderomotive force predicted by \cite{Laming:2004,Laming:2015} is expected to act on the plasma in the chromosphere, a region historically understudied, but the subject of intense investigation since the launch of the Interface Region Imaging Spectrometer \cite[IRIS;][]{depontieu:2014} spacecraft in 2013. The region of the spectrum observed by IRIS includes two C~II lines (at 1334 \& 1335~\AA), two Si~IV lines (at 1394 \& 1403~\AA), and the two Mg~II k \& h lines (2796 \& 2803~\AA\ respectively). The C~II and Si~IV lines observed by IRIS probe the \cref{upper chromosphere and transition region \cite[cf.][]{Rathore:2015,Testa:2023,Sainzdalda:2024}}, and as such provide a good overview of the plasma diagnostics in \cref{these locations}. It has been suggested by \cite{MartinezSykora:2023} that Si~IV should exhibit increased non-thermal velocities implying waves associated with the ponderomotive force, with \cite{Long:2024} finding hints of this behaviour in \cref{the leading and following parts of} an active region traversing the solar disk. The two Mg~II h \& k lines observed by IRIS probe optically thick plasma in the solar chromosphere, and as such can be used to study velocity gradients through the chromosphere \cite{Pereira:2013}. They can also be inverted to derive physical parameters such as micro-turbulence velocity, electron density, temperature, and line-of-sight velocity \cite[e.g., using the IRIS$^2$ approach developed by][]{Sainzdalda:2019}. This technique was used by \cite{Testa:2023} to identify hints of increased chromospheric micro-turbulence velocity associated with increased FIP bias in the solar corona. 

\cref{Here, we use a full-disk mosaic produced by the IRIS spacecraft to probe the interface between the solar chromosphere and transition region and try to identify signatures of the ponderomotive force predicted by \cite{Laming:2004} and \cite{MartinezSykora:2023} in chromospheric plasma. Building on previous work by \cite{Long:2024} and \cite{Testa:2023} who tracked a single active region for several days, we use the full-disk mosaic to study multiple active regions at different evolutionary stages using each of the C~II, Si~IV, and Mg~II lines observed by IRIS simultaneously. We can also use the classification by \cite{Mihailescu:2022} of the active regions studied here the characterise the properties of the leading and following polarities of the individual active regions, which} is important given the asymmetry in flux density and previous results identifying distinct differences in wave behaviour between the leading and following polarities of active regions \citep[cf.][]{Giannattasio:2013}.

\section{Observations}

\begin{figure}
    \centering
    \includegraphics[width=1\linewidth]{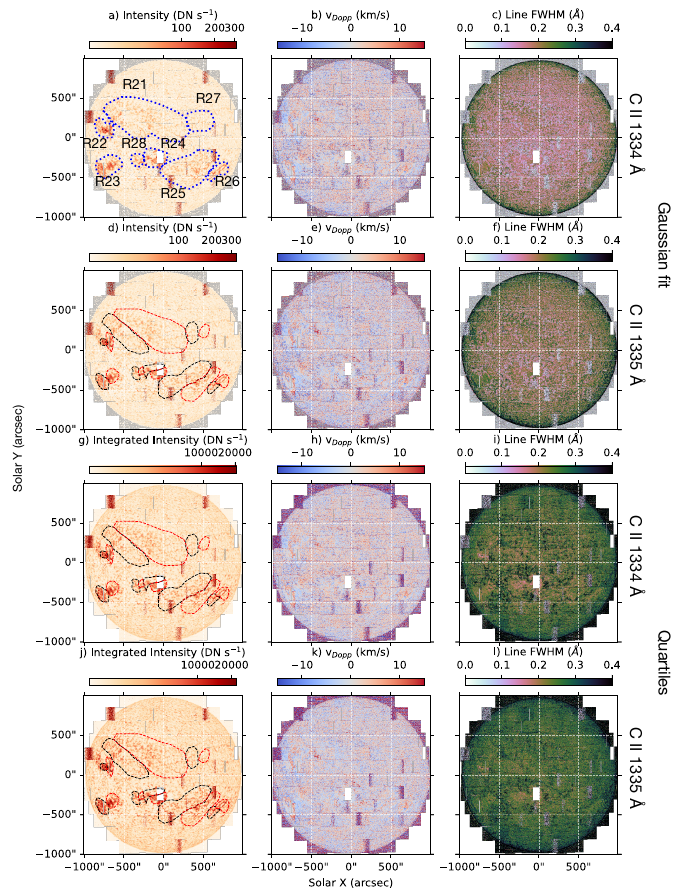}
    \caption{Full disk mosaics taken by the IRIS spacecraft in the \cref{C~II~1334~\AA\ and C~II~1335~\AA} spectral lines\cref{, with the lines indicated by the text on the right hand side of each row. The upper two rows show the lines as} fitted using a single Gaussian\cref{, while the bottom two rows show the results derived using a quartile approach (see text for details). In each case,} the left column \cref{shows} the fitted peak intensity, the centre column the Doppler velocity, and the right column the \cref{full width at half maximum (FWHM) of the line (to enable a direct comparison between the Gaussian fitting and quartile techniques)}. Panel~a shows the active regions defined by \cite{Mihailescu:2022} and analysed here, with \cref{panels~d, g, and j} showing the leading (red) and following (black) polarities of the corresponding active regions.}
    \label{fig:cii_context}
\end{figure}

\begin{figure}
    \centering
    \includegraphics[width=0.9\linewidth]{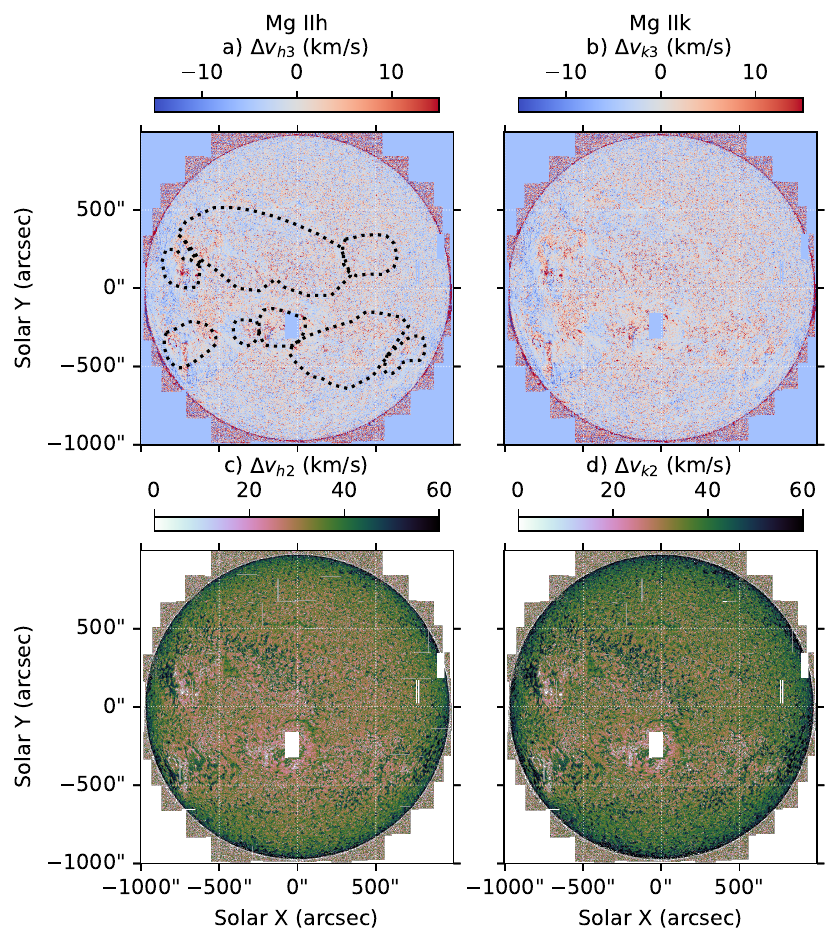}
    \caption{Full disk mosaics in the Mg~II h \& k lines showing the derived h/k 3 velocity (top row) and h/k 2 separation (bottom row) for the Mg~II h (left column) and k (right column) lines. The regions of interest have been indicated on panel (a) as in \cref{Figures~\ref{fig:siiv_context}a and \ref{fig:cii_context}a.}}
    \label{fig:MgII_features_context}
\end{figure}

The IRIS spacecraft regularly takes full disk mosaics of the solar disk using a series of pointings that take up to $\sim$18~hours per observation. Each full disk mosaic consists of $\sim$185 individual pointings of the spacecraft, \cref{using} a 64-step raster with 2'' steps taken at each pointing, enabling a region of $\sim$180''$\times$128'' to be observed in each case across regions of the spectrum covering the C~II~1334 \& 1335~\AA, Si~IV~1394 \& 1403~\AA, and Mg~II~h \& k spectral lines. Following observation, the individual pointings are processed using the IRIS data processing pipeline and stitched together to produce individual 3-dimensional data cubes (X, Y, $\lambda$) for each spectral window. These data cubes are then made available online via the IRIS website\footnote{\url{https://iris.lmsal.com/mosaic_index.html}}. Here, we focus on a full disk mosaic taken by the IRIS spacecraft on 18~October~2015. This IRIS mosaic was contemporaneous with a full disk mosaic taken by the \emph{Hinode} spacecraft which has been previously studied in detail by \cite{Mihailescu:2022}, and which is the focus of work elsewhere in this special issue by \cite{Spruksta:2025}.

The full disk mosaic raster files for all of the spectroheliograms taken during the \cref{18~October~2015} observation were first downloaded from the IRIS website. The Si~IV~1394~\AA\ \cref{and} Si~IV~1403~\AA\ spectral lines were each fitted using single Gaussian fits to derive the peak intensity, Doppler velocity, and nonthermal velocity \cref{(by first converting the Gaussian width to full width at half maximum), with the resulting plots} shown in Figure~\ref{fig:siiv_context}. This is a similar approach to that previously used by \cite{Long:2024}.

\begin{figure}
    \centering
    \includegraphics[width=1\linewidth]{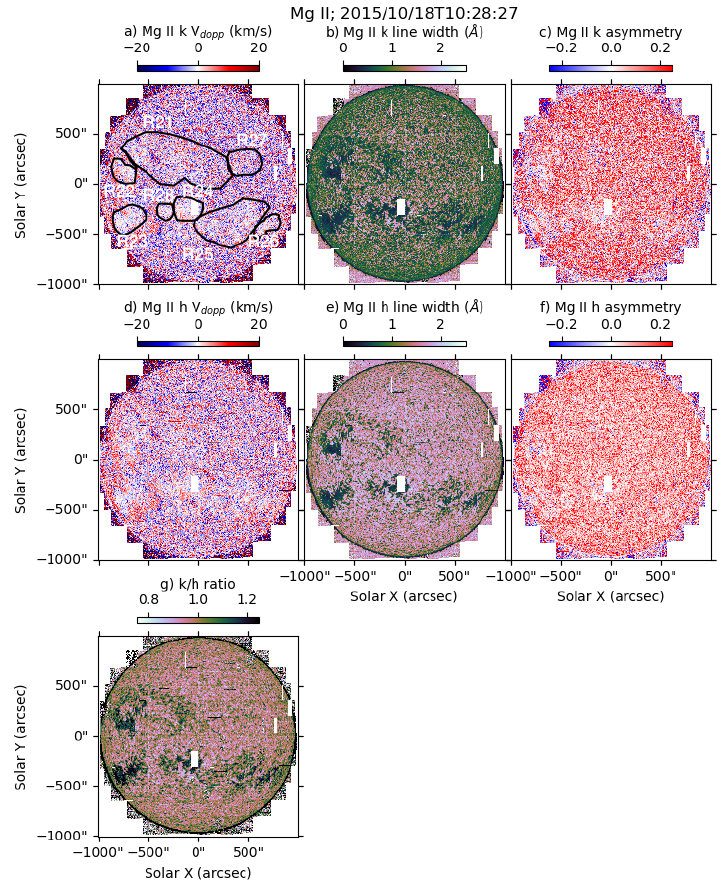}
    \caption{Full disk mosaics showing the Doppler velocity (panels a \& d), line width (panels b \& e), and line asymmetry (panels c \& f) for the Mg~II k (top row; panels a, b, c) and Mg~II h (bottom row; panels d, e, f) lines. Panel~g shows the ratio of integrated intensity across the k and h lines. Panel~a also shows the different active regions analysed here for context.}
    \label{fig:MgII_plasma_params}
\end{figure}

In contrast to the two Si~IV lines, which are both relatively clean lines produced by optically thin plasma and therefore easily fitted using single Gaussians, the \cref{C~II and} Mg~II h \& k lines are produced by optically thick plasma, and as a result are much more difficult to fit and interpret \cref{\cite[cf.][]{Rathore:2015}.} As the full disk mosaic examined here contained a variety of different regions of the solar atmosphere including active regions, quiet Sun, and coronal holes, each with varying signal to noise ratios (thus making identification of \cref{optically thick} lines more difficult) we used two different techniques to identify and characterise the\cref{m. First, previous work \cite[e.g.,][]{Ghosh:2017} suggests that the C~II lines can be well represented using single Gaussians in certain locations, so this approach was initially taken with the C~II lines. In addition}, following the lead of \cite{Long:2024}, the Mg~II h \& k lines were fitted using a Python implementation of the IDL iris\_get\_mg\_features\_lev2.pro routine. This approach \cref{identifies} the k1 (h1), k2 (h2), and k3 (h3) points of the k (h) line\cref{,} thus allowing an estimation to be made of velocity gradients in different parts of the solar chromosphere \cite[see e.g.,][for more details]{Pereira:2013,Long:2024}. The results of this approach are shown in \cref{the top two rows of} Figure\cref{~\ref{fig:cii_context} for Gaussian fits to the the C~II lines, as well as in Figure~}\ref{fig:MgII_features_context} for the h/k~3 velocity ($\Delta v_{h/k 3}$) and the h/k~2 separation ($\Delta v_{h/k 2}$), corresponding to the velocity gradients in the upper- and mid-chromosphere respectively.

The second \cref{analysis} approach followed the lead of \cite{Kerr:2015} and used a non-parametric quartile approach to characterise the \cref{C~II and} Mg~II h \& k lines. This approach makes no assumption about the shape of the corresponding spectral line, instead using normalised cumulative distribution functions of intensity vs.\ wavelength for each line and estimating the 25\% (Q$_1$), 50\% (Q$_2$), and 75\% (Q$_3$) quartiles. These quartiles can then be used to estimate,
\begin{enumerate}
    \item $\lambda_c = Q_2$; the wavelength of the line centroid and hence Doppler velocity given a rest wavelength,
    \item $W = Q_3 - Q_1$; the line width, and \\
    \item A = $\frac{(Q_3-Q_2)-(Q_2-Q_1)}{Q_3-Q_1}$; the line asymmetry.
\end{enumerate}
The resulting full disk mosaic images are shown in \cref{the bottom two rows of} Figure\cref{~\ref{fig:cii_context} for the C~II lines and Figure~}\ref{fig:MgII_plasma_params} for the \cref{Mg~II} k (top row) and h (bottom row) lines respectively. Figure~\ref{fig:MgII_plasma_params}d also shows the ratio of integrated intensity across the k and h lines, providing an insight into the opacity of the plasma \cite[cf.][]{Schmeltz:1997,Mathioudakis:1999}.

\section{Results}

The region of the solar atmosphere probed by the IRIS spacecraft is vitally important for understanding how mass and energy evolve through the solar atmosphere as it covers the region where the plasma-$\beta = 1$ and hence where mode conversion and reflection/refraction of waves can occur. The full disk mosaics produced by the IRIS spacecraft are an underused resource that enable long term tracking of different features but also a direct, contemporaneous comparison between active regions of different ages and evolutionary stages. Previous work by \cite{Mihailescu:2022} has shown variations in coronal measurements of FIP bias distributions between active regions of different evolutionary stages, and the dataset examined here offers an opportunity to probe chromospheric signatures associated with those same active regions.

\begin{figure}
    \centering
    \includegraphics[width=1\linewidth]{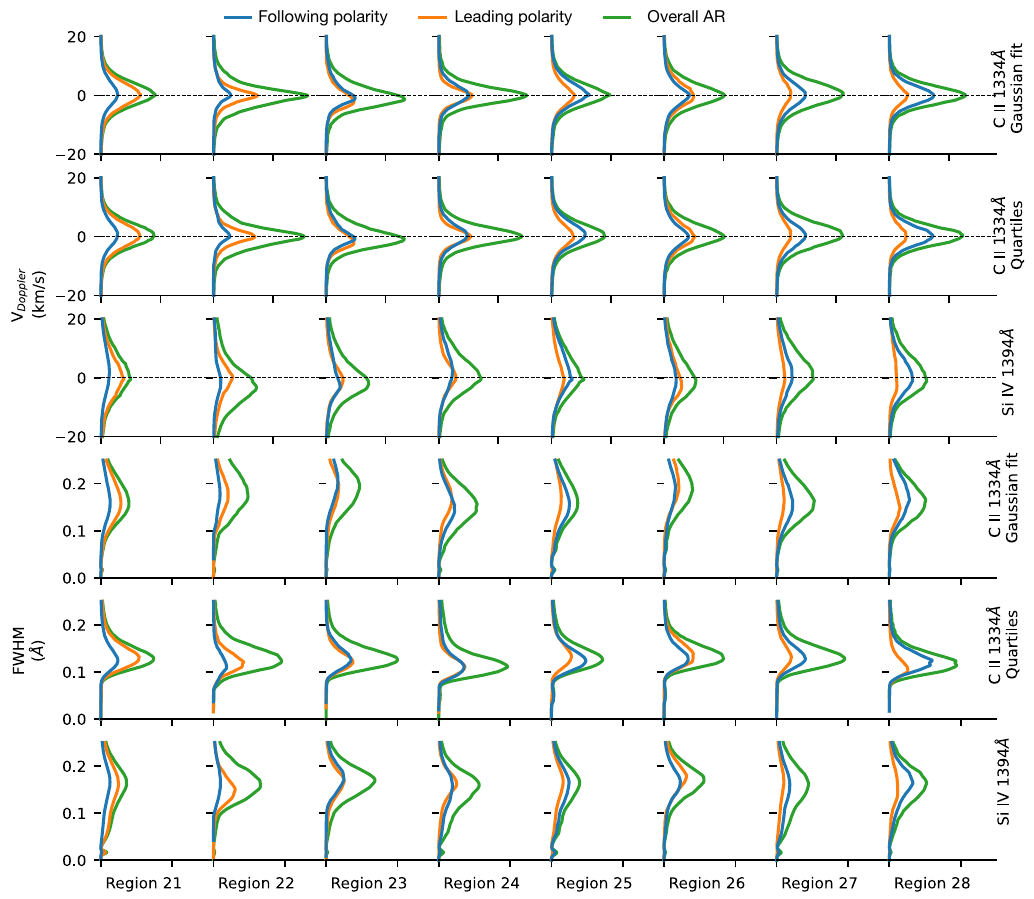}
    \caption{Kernel density estimation (KDE) plots showing the variability in Doppler velocity (upper \cref{three} rows) and \cref{FWHM} (bottom \cref{three} rows) for the C~II~1334~\AA\ \cref{(assuming both a Gaussian fit and quartile approach)} and Si~IV~1394~\AA\ lines\cref{. In each case, the wavelength is indicated by the labels on the right hand side, with the regions of interest shown by the label on the bottom row. T}he Doppler velocity distributions are centred on zero, with the \cref{FWHM} mostly distributed about \cref{0.1 - 0.2~\AA\ regardless of technique used to derive it. Note that the C~II~1335~\AA\ and Si~IV~1403~\AA\ lines display comparable behaviour to the C~II~1334~\AA\ and Si~IV~1394~\AA\ lines respectively and are not shown here for clarity.}}
    \label{fig:siiv_cii_kde}
\end{figure}

As previously noted by \cite{Long:2024}, the different plasma parameters probed by IRIS exhibit distributions of values in different regions, making it difficult to interpret them using single representative values such as mean or median. Instead, Kernel Density Estimation (KDE) plots were used to visualise the individual distributions and directly compare how they vary between different parts of the active regions studied. Figure~\ref{fig:siiv_cii_kde} shows the results of this analysis for the Si~IV~1394~\AA\ and C~II~1334~\AA\ spectral lines, focussing in particular on the Doppler velocity (top \cref{three} rows) and \cref{full width at half maximum (FWHM;} bottom \cref{three} rows). In each case, the distributions of values for the leading (orange) and following (blue) polarities along with the overall active region (green) are shown for each spectral line. \cref{The C~II~1335~\AA\ and Si~IV~1403~\AA\ lines display comparable behaviour to the C~II~1334~\AA\ and Si~IV~1394~\AA\ lines respectively for each region, and are not shown here for clarity. In addition, the Doppler velocity and FWHM for the C~II~1335~\AA\ line were calculated using both the fitted Gaussian and quartile approach, with the technique used noted on the right-hand side of each row.}

The Doppler velocity in each case is quite evenly distributed, centred on zero with a slight skewness visible for \cref{Si~IV~1394~\AA\ line} in regions 22 \& 23 in particular. \cref{The Si~IV~1394~\AA\ line also shows a much broader distribution of Doppler velocity values compared with the C~II~1334~\AA\ line (regardless of fitting technique).} However, there are no clear discrepancies between the leading and following polarities of the active regions \cref{for any of the spectral lines}, suggesting no obvious dependence on the evolutionary stage of the active region. For the \cref{FWHM, the distributions for a given spectral line tend to show no clear variation between the different regions. However, the C~II~1334~\AA\ distributions show a slight difference between analysis technique, with the Gaussian fit distributions tending to be broader and peaking closer to 0.2~\AA. In contrast, the quartile distributions tend to be much narrower and peaking closer to 0.1~\AA. The Si~IV~1394~\AA\ distributions (also derived using a Gaussian fit) are comparable to the C~II~1334~\AA\ distributions. It is noticeable that there is some evidence of increased FWHM in the leading (regions 21, 22, and 26) or following (regions 25, 27, and 28) polarities, with regions 23 \& 24 showing no variation between polarities.} 

\begin{figure}
    \centering
    \includegraphics[width=1\linewidth]{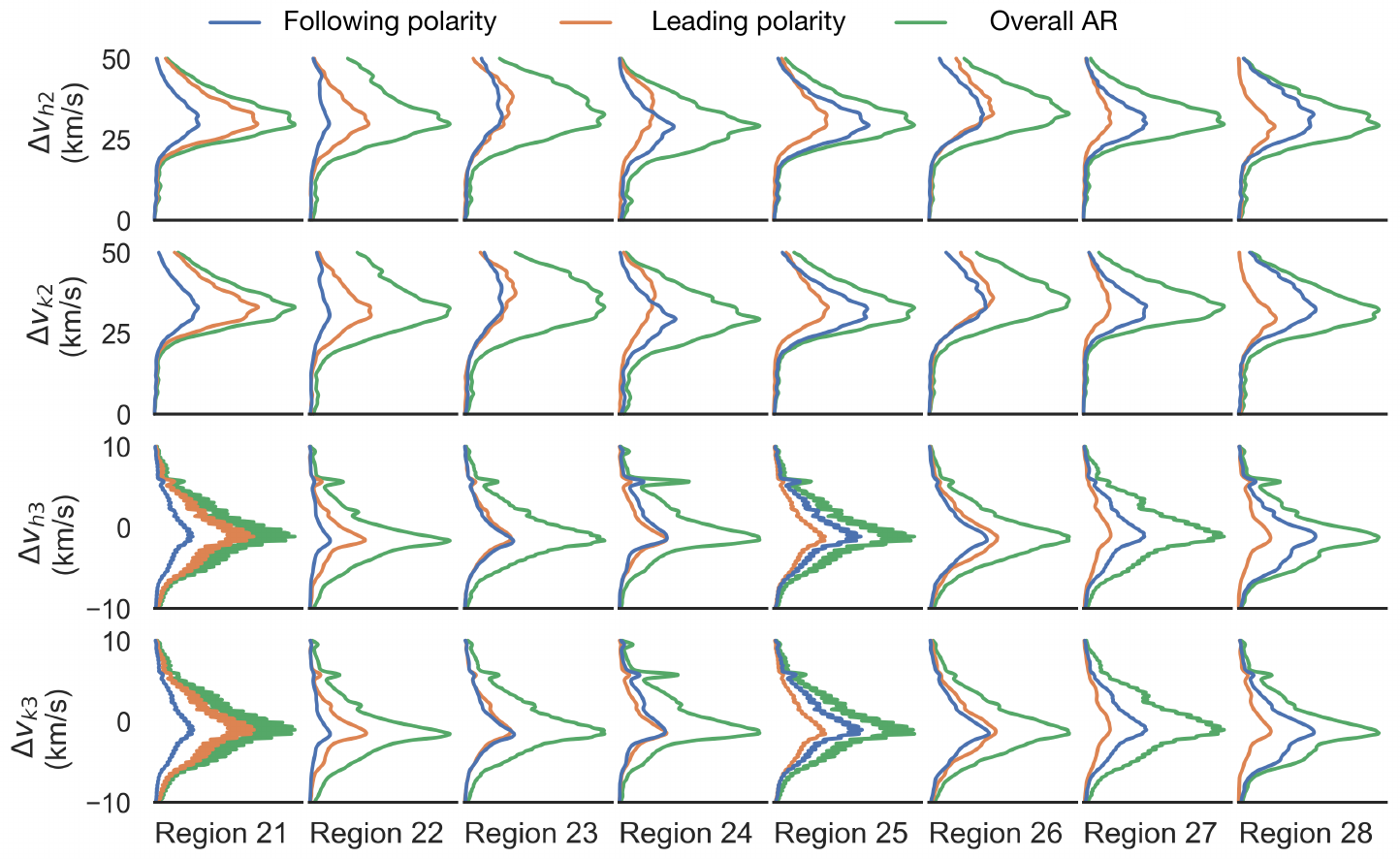}
    \caption{KDE plots showing the variability in chromospheric velocity gradients derived using the Mg~II h \& k lines. Properties are noted on the left hand side of the figure for the lines given on the right hand side, with the regions noted along the bottom axis.}
    \label{fig:mgii_plasma_kde}
\end{figure}

In contrast to the C~II and Si~IV lines, which probe the upper chromosphere and transition region, the Mg~II h \& k lines probe different parts of the chromosphere defined by the behaviour of different parts of the optically thick Mg~II lines. The velocity gradients of the mid- and upper-chromosphere were first probed by examining the h/k 2 separation ($\Delta v_{h/k 2}$) and h/k 3 velocity ($\Delta v_{h/k 3}$) respectively. The corresponding KDE distributions are shown in Figure~\ref{fig:mgii_plasma_kde}. As expected, the different parameters given for the h \& k lines are broadly comparable, with the plasma probed by the h \& k lines differing in height by several 10's of km in the chromosphere \citep[cf.][]{Leenaarts:2013}. In general, the h2 \& k2 parameters probing the velocity gradient in the mid-chromosphere tend to peak above 25~km/s, with some small differences between leading and following polarity regions in the different active regions. However, the velocity gradient in the upper chromosphere (shown by the h3 \& k3 parameters) tend to be distributed about 0, with both positive and negative velocities observed, and again slight differences between the leading and following polarities in different active regions.

\begin{figure}
    \centering
    \includegraphics[width=1\linewidth]{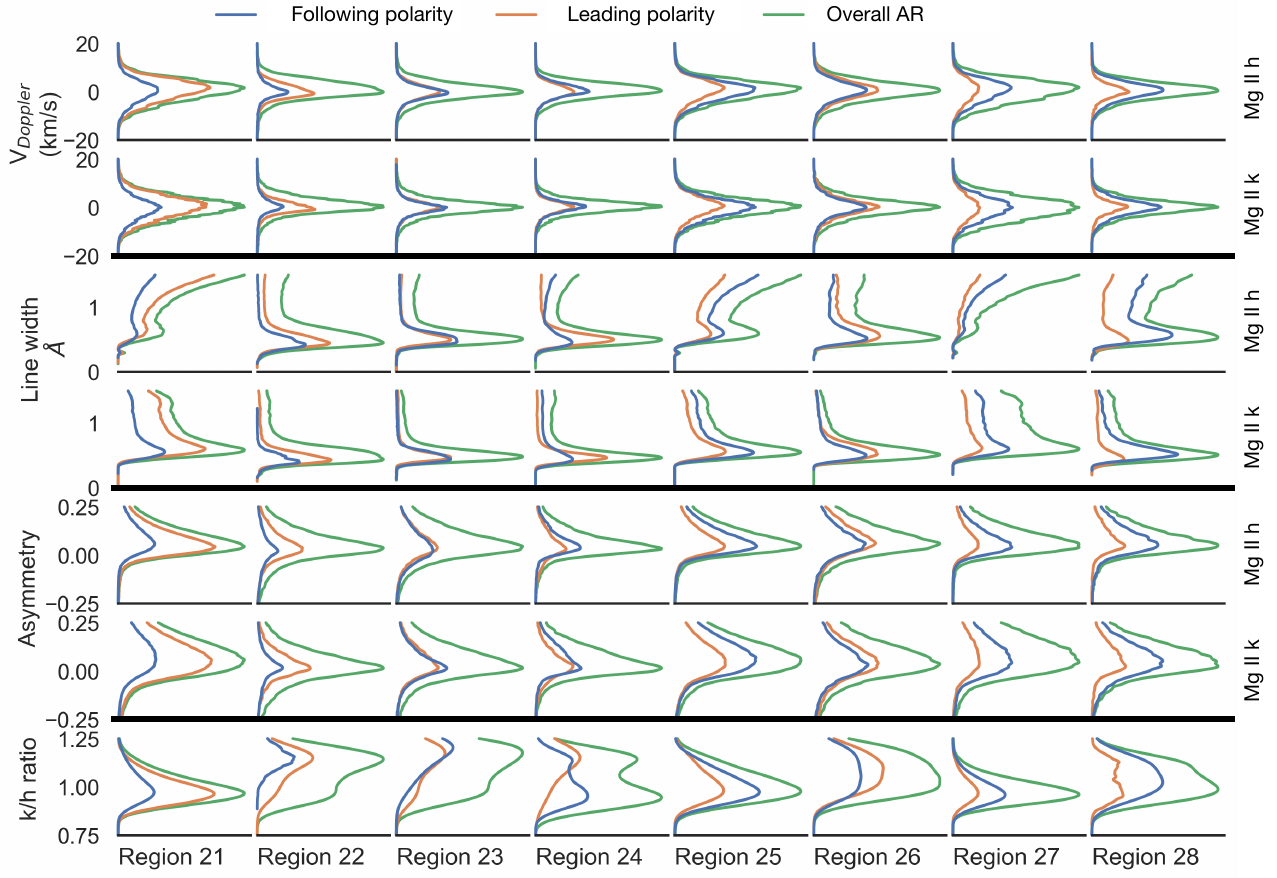}
    \caption{KDE plots showing the variability in plasma properties of the Mg~II h \& k lines derived using the quartile method as outlined in the text. Properties are noted on the left hand side of the figure, with the regions noted along the bottom axis.}
    \label{fig:mgii_plasma_param_kde}
\end{figure}

Given the low signal to noise and the corresponding difficulty in fitting multiple Gaussians to the Mg~II h \& k spectral lines in some pixels using the iris\_get\_mg\_features\_lev2 approach, the Mg~II lines were also analysed using a quartile analysis approach. This enabled the Doppler velocity, width, and asymmetry of the h \& k lines to be estimated, with the corresponding KDE plots shown in \cref{Figure~\ref{fig:mgii_plasma_param_kde}}. As expected, the distributions of Doppler velocity are mostly centred on 0 for each active region, indicating a combination of both upflows and downflows in each active region. The line width distributions each tend to have a peak around 0.5~\AA\, with the strong increase above 1~\AA\ seen in some cases (e.g., Regions 21, 25, 27, 28) consistent with noisy data or low signal to noise. The distributions of line asymmetry are mostly positive, indicating a greater enhancement of the line redward of the line core vs. blueward, consistent with previous observations of a flaring active region made by \cite{Kerr:2015}. 

In addition to these plasma parameters, the quartile analysis also enables a comparison to be made between the integrated intensity of the k \& h lines, offering an insight into the opacity of the line. Depending on the region of the sun being observed, previous work has indicated a typical value in the range of R$_{kh}\sim 1.14-1.46$ \cite{Kerr:2015,Lemaire:1981}. The bottom row of Figure~\ref{fig:mgii_plasma_kde} shows the distribution of k/h values for the different active regions studied here. It is clear that there is a wide range of behaviours in the different regions, with e.g., Region~21 showing a simple single-peaked distribution peaking around 1 for both leading and following polarities, while for Region~24 the leading polarity region peaks around 1.2 with the following polarity region peaking around 0.9. This behaviour indicates variability in whether the \cref{plasma} is optically thick or optically thin in different parts of the active regions studied\cref{, with implications for e.g., the damping, scattering, and propagation of any waves propagating through that region of the atmosphere.}

\section{Discussion}

\cref{The full disk mosaics taken by the IRIS spacecraft are a unique and underexplored resoucre which provide a snapshot of active regions at different evolutionary stages observable on the solar disk at the time of the observation.} Previous work by \cite{Mihailescu:2022} used \cref{a} contemporaneous full disk mosaic taken by \emph{Hinode}/EIS \cref{to classify individual active regions at different evolutionary stages and examine their associated FIP bias}, finding a weak correlation between FIP effect and evolutionary stage of the active region. Here we have \cref{built on the work of \cite{Mihailescu:2022}} to \cref{examine} the chromosphere and transition region \cref{using} the C~II, Si~IV, and Mg~II lines observed by the IRIS spacecraft\cref{, and focusing on identifying a relationship between the plasma diagnostics in this part of the atmosphere and FIP bias rather than deriving FIP bias directly.} 

As described in Table~1 of \cite{Mihailescu:2022} the active regions studied here are predominantly simple bipolar regions, with a single active region nest which has by this time evolved into a filament channel. \cref{In general, the Doppler velocities derived for the C~II and Si~IV spectral lines are quite symmetric about zero, as shown in Figure~\ref{fig:siiv_cii_kde}. This is consistent with no large bulk plasma motion in the active regions, with the leading vs. following polarities only showing slight differences in the number of pixels studied. In contrast, six of the eight regions studied show clear differences in FWHM between the leading and following polarities of the different active regions. This is interesting, and suggests differences in unresolved wave behaviour in the different polarities. Regions 21, 22, and 26 show increased FWHM in the leading regions, but have no clear similarities, with region 21 corresponding to a large decayed filament channel, region 22 corresponding to a named active region with spots, while region 26 is a dispersed active region. In contrast, regions 25, 27, and 28 show increased FWHM in the following polarities, and correspond to simple, dispersed active regions. Regions 23 and 24 show no variation between polarities and both correspond to compact active regions with spots. The C~II and Si~IV lines} predominantly probe the upper chromosphere and transition region, high above the location where the ponderomotive force is predicted to act. Previous work has hinted at some identifiable variability, \cref{with these observations suggesting that more work is required to identify and predict the spectral signatures expected for the waves driving the ponderomotive force in this part of the atmosphere.} 

The distributions corresponding to the two Mg~II lines analysed using the different techniques are potentially more interesting. As optically thick lines, both the Mg~II h \& k lines can be used to probe different parts of the solar atmosphere simultaneously, and can be strongly affected by variations in plasma properties such as density and temperature. Differences in the h/k~2 and h/k~3 line locations provide an insight into velocity gradients at different heights in the solar chromosphere, with the result that they have previously been used to try and identify signatures of the ponderomotive force induced by waves coming from the corona. The KDE distributions shown in Figure~\ref{fig:mgii_plasma_kde} show distinct differences between the leading and following polarities of some of the different active regions studied. Although regions 21, 22, and 27 are all located in the northern hemisphere, region 27 has a much more pronounced following polarity distribution, while the following polarity distribution is negligible in region 22 (most likely due to its proximity to the solar limb), with region 21 exhibiting a strongly peaked leading polarity distribution in each parameter. In the southern hemisphere, regions 23 and 24 have no clear difference between the following and leading polarity distributions, region 26 has comparable distributions between both polarities, while regions 25 and 28 both have strongly peaked following polarity distributions.

However, the distributions of $\Delta v_{h/k 3}$ illustrate the issues with using the Python implementation of the iris\_get\_mg\_features\_lev2 algorithm, with much more noise identifiable in the distributions most likely a result of the lower number of useable pixels due to the increased number of parameters required when using this approach to fit the lines. To overcome this issue a quartile approach was also used to study the Mg~II lines. Although the KDE distributions shown in Figure~\ref{fig:mgii_plasma_param_kde} do not suffer from this issue, the effects of noise can be seen in the strongly peaked tail in the distributions of line width, primarily for the Mg~II~h line (3rd row from the top of Figure~\ref{fig:mgii_plasma_param_kde}. However, this approach also provides some interesting insights into the state of the plasma. 

In each case, the Doppler velocity is peaked at or close to 0, with some very slight differences between the leading and following polarities most likely due to position on disk. The line width is also relatively uninformative; both lines tend to have peaks in the distributions of line width at approximately 0.4-0.6~\AA, with some of the distributions showing increases above 1~\AA\ most likely due to poor signal to noise. However, the line asymmetry does exhibit some noticeable differences between the leading and following polarities. Although each distribution is predominantly skewed positive, there are some variations between the leading and following polarities. The line asymmetry can be interpreted as due to a combination of both red- and blue-shifted plasma motions along the line of sight within a single pixel \cite[cf.][]{Kerr:2015}, but it has also been suggested by \cite{Heinzel:1994} that increased blue asymmetry could be the result of downward propagating plasma absorbing radiation from the red side of the line core. However, this suggestion was made based on observations of plasma evolution during a flare rather than from an active region as with here, so may not be appropriate in this case; modelling is required to confirm the behaviour described here.

The parameter which exhibits the most \cref{interesting} variability is the k/h ratio shown in the bottom row of Figure~\ref{fig:mgii_plasma_param_kde}. The k/h ratio is expected to be less than 2 if the plasma is optically thick, with previous work finding a value in the range $1.14--1.46$ \cite{Lemaire:1981,Lemaire:1984,Kerr:2015}. As shown in Figure~\ref{fig:mgii_plasma_param_kde}, the k/h ratio varies strongly between the different regions, with regions 21, 25, 27 (generally corresponding to large, well-dispersed, decayed active regions) exhibiting relatively narrow single peaked distributions peaking around $k/h=1$. The distributions corresponding to regions 26 \& 28 (which are much smaller but also quite decayed) are much broader, but also single peaked around $k/h=1$. However, the distributions for regions 22, 23, and 24 are double peaked, with peaks either side of $k/h=1$. These regions contain much stronger magnetic field (and in fact are the only regions in the dataset studied here which contain spots), indicating that the combination of increased magnetic field strength in the active region core and weaker magnetic field strength towards the edges affects the distribution of opacity of the observed plasma. \cref{Regions 24 and 28 also show differences between the distributions for the leading and following polarities, suggesting plasma properties which may affect the behaviour of waves propagating through this region of the atmosphere.}

\section{Conclusions}

The aim of this project was to compare a wide variety of chromospheric plasma diagnostics derived using spectral lines observed by the IRIS spacecraft to quantify any signatures associated with coronal FIP bias. Of the spectral lines observed by IRIS, the C~II and Si~IV lines probe the upper chromosphere and transition region, whereas current theories for explaining observed FIP bias invoke a ponderomotive force acting in the upper photosphere and low to mid chromosphere. \cref{Previous work has suggested that the width of the Si~IV spectral lines, and hence nonthermal velocity, should exhibit behaviour consistent with unresolved waves related to the fractionation process \cite[cf.][]{MartinezSykora:2023,Long:2024}. Although not very clear from the different active regions studied here, there are differences between the distributions of FWHM for the leading and following polarities of the individual active regions. These active regions were previously identified and studied in the corona by \cite{Mihailescu:2022}, who found a slight difference between the FIP bias of the leading and following polarities of the studied active regions. However, the differences identified using the C~II and Si~IV lines here are quite small, suggesting} either that the signatures \cref{of the fractionation process} are quite faint and require a stronger magnetic field to become more apparent (as the regions studied here are generally quite weak and dispersed compared to those previously studied), or there is no \cref{strong} relationship. A full and complete understanding of this process and conclusive determination of whether it can be identified using the Si~IV or C~II lines observed by IRIS \cref{therefore requires} validation using modelling\cref{, which we intend to do in future work}.

In contrast to the Si~IV and C~II lines, the Mg~II lines observed by IRIS are optically thick and probe the region lower down in the solar chromosphere closer to where the ponderomotive force is predicted to fractionate the plasma. Previous work using the iris\_get\_mg\_features\_lev2 algorithm developed by \cite{Pereira:2013} has identified no clear signatures of this process \cite[cf.][]{Long:2024}. However, this algorithm relies on fitting multiple functions simultaneously to derive the different parameters of the Mg~II h \& k lines, and as a result can be affected by low signal to noise in quieter regions. In contrast, a quartile approach to deriving the properties of the Mg~II h \& k lines makes no assumptions about shape or intensity of the lines \cite[cf.][]{Kerr:2015}, making it more robust and better able to identify properties such as Doppler velocity, line width, and line asymmetry. However, using this approach revealed no significant deviations between the different regions studied in Doppler velocity, line width, or line asymmetry.

The Mg~II h \& k lines are produced by transitions to a common lower energy level from finely split upper levels and as a result can be used to estimate the opacity of the plasma by comparing the ratio of the line intensities. Using this approach, significant differences could be identified between the different regions studied. In particular, some regions exhibited narrow single peak distributions, some had broad single peaked distributions, while 3 regions had double peaked distributions. It should be noted that the regions with double peaks in the distributions of k/h ratio also had the highest average FIP bias of these regions as estimated by \cite{Mihailescu:2022}. These differences in the k/h ratio of the different regions are particularly interesting given that the opacity of a plasma can play a role in the propagation of waves in that plasma. \cref{Subsequent work by \cite{Mihailescu:2023} identified the importance of resonant vs.\ non-resonant waves in driving fractionation of the plasma in different magnetic loop populations. Although this was shown using modelling, the effects of plasma opacity on this process and implications of how variations in the height at which this fractionation occurs affect the observed spectral signatures of the plasma remain to be confirmed. It should also be noted that the waves driving the fractionation process exhibit strong temporal variation, whereas the observations analysed here provide a spatial analysis of active regions at different evolutionary stages, but with no temporal resolution. It may therefore be that identifying signatures of the fractionation process require much higher temporal cadence observations of e.g., $<$10~s, although exact numbers will require modelling of the fractionation process and associated spectral signatures. Nonetheless, the work described here provides a useful constraint on predictions of previously underexplored spectral lines, and we look forward to investigating this more thoroughly using both higher cadence observations and by combining observations with modelling in future work.}

\dataccess{No new data were generated as part of this study. The IRIS full disk mosaics can be downloaded from the IRIS website (\href{https://iris.lmsal.com/mosaic_index.html}{https://iris.lmsal.com/mosaic\_index.html})}

\aucontribute{DML conceived the study. EP carried out the data reduction and scientific analysis with assistance from DML. TM assisted with identification and analysis of the different active regions. LAH assisted with analysis of the IRIS data. DML drafted the manuscript with help from EP. All authors read and approved the manuscript.}

\competing{The author(s) declare that they have no competing interests.}

\funding{LAH is supported by a Royal Society-Research Ireland University Research Fellowship (URF$\backslash$R1$\backslash$241775)
}

\ack{
\cref{The authors wish to thank the two referees whose suggestions helped to improve the paper.}
We thank the Royal Society for its support throughout the organization of the Theo Murphy meeting on "Solar Abundances in Space and Time" and production of this special issue.
IRIS is a NASA small explorer mission developed and operated by LMSAL with mission operations executed at NASA Ames Research Center and major contributions to downlink communications funded by ESA and the Norwegian Space Centre.
This research used the following software packages during analysis of the data and preparation of this paper: Numpy \citep{Harris:2020}, Scipy \cite{Virtanen:2020}, SunPy \citep{Sunpy:2020}, Seaborn \citep{Waskom:2021}, Matplotlib \citep{Hunter:2007}, Astropy \citep{Astropy:2022}.
}


\bibliographystyle{RS} 
\bibliography{bibliography} 

\end{document}